\documentstyle[12pt,epsf,epsfig]{article}
\def\beq{\begin{equation}}
\def\eeq{\end{equation}}
\def\bea{\begin{eqnarray}}
\def\eea{\end{eqnarray}}
\def\bq{\begin{quote}}
\def\eq{\end{quote}}
\def\ie{{\em i.e.}}

\def\vev#1{\langle #1\rangle}

\def\gappeq{\mathrel{\rlap {\raise.5ex\hbox{$>$}}
{\lower.5ex\hbox{$\sim$}}}}

\def\lappeq{\mathrel{\rlap{\raise.5ex\hbox{$<$}}
{\lower.5ex\hbox{$\sim$}}}}

\def\Toprel#1\over#2{\mathrel{\mathop{#2}\limits^{#1}}}

\begin{document}
\pagestyle{empty}
\begin{flushright}
{ACT-04/02} \\
{CERN-TH/2002-117} \\
{CTP-11/02} \\
hep-ph/0205336 \\
\end{flushright}
\vspace*{5mm}
\begin{center}
{\large {\bf FLIPPING $SU(5)$ OUT OF TROUBLE}}
\\
\vspace*{5mm}
{\bf John Ellis}$^1$,
{\bf D.V. Nanopoulos}$^{2,3,4}$ and
{\bf J. Walker}$^2$\\ \vspace{0.5cm}
$^1$ Theoretical Physics Division, CERN,\\
CH 1211 Geneva 23, Switzerland\\ \vspace{0.5cm}
$^2$ Center for Theoretical Physics, Department of Physics, 
Texas A\&M
University,\\ College Station, TX 77843--4242, USA\\ \vspace{0.5cm}
$^3$ Astroparticle Physics Group, Houston Advanced Research 
Center (HARC),\\
The Mitchell Campus, The Woodlands, TX 77381, USA\\ \vspace{0.5cm}
$^4$ Chair of Theoretical Physics,
Academy of Athens,
Division of Natural Sciences,\\
28~Panepistimiou Avenue,
Athens 10679, Greece\\

\vspace*{5mm}

{\bf ABSTRACT} \\ \end{center}

\vspace*{5mm}

\noindent
Minimal supersymmetric $SU(5)$ GUTs are being squeezed by the recent
values of $\alpha_s$, $\sin^2 \theta_W$, the lower limit on the lifetime
for $p \to {\bar \nu} K$ decay, and other experimental data. We show how
the minimal flipped $SU(5)$ GUT survives these perils, accommodating the
experimental values of $\alpha_s$ and $\sin^2 \theta_W$ and other
constraints, while yielding a $p \to e/\mu^+ \pi^0$ lifetime beyond the
present experimental limit but potentially accessible to a further round
of experiments. We exemplify our analysis using a set of benchmark
supersymmetric scenarios proposed recently in a constrained MSSM
framework.

\vspace*{1cm}

\begin{flushleft} 
CERN-TH/2002-117 \\
May 2002
\end{flushleft}

\newpage


\setcounter{page}{1}
\pagestyle{plain}

One of the key pieces of circumstantial evidence in favour of grand 
unification has long been the consistency of the gauge couplings measured 
at low energies with a common value at some very high energy scale, once 
renormalization effects are taken into account. This consistency is 
significantly improved when light supersymmetric particles are included in 
the renormalization-group running, in which case the agreement improves 
to the per-mille level~\cite{EKN}.

However, this circumstantial evidence is not universally accepted as
convincing. For example, it has recently been suggested that the
logarithmic unification of the gauge couplings is as fortuitous as the
apparent similarity in the sizes of the sun and moon~\cite{Ibanez}.
Alternatively, it has been argued that the unification scale could be as
low as 1 TeV, either as a result of power-law running of the effective
gauge couplings in theories with more than four dimensions~\cite{power},
or in theories with many copies of the $SU(3) \times SU(2) \times U(1)$
gauge group in four dimensions~\cite{Georgi}.

For some time now, detailed calculations have served to
emphasize~\cite{Ross} how much fine tuning is needed in models with
power-law running to reproduce the effortless success of supersymmetric
grand unification with logarithmic running of the gauge couplings.
Moreover, data from particle physics and cosmology provide independent
hints for low-energy supersymmetry. Precision electroweak data favour
quite strongly a low-mass Higgs boson~\cite{LEPEWWG}, as required in the
minimal supersymmetric extension of the Standard Model
(MSSM)~\cite{SusyHiggs}, and the lightest supersymmetric particle is a
perfect candidate~\cite{EHNOS} for the cold dark matter thought by
astrophysicists to infest the Universe. Many studies have shown
that these and other low-energy data - such those on $b \rightarrow s
\gamma$ decay~\cite{bsg} and $g_\mu - 2$~\cite{g-2} - are completely
consistent with low-energy supersymmetry, and a number of benchmark
supersymmetric scenarios have been proposed~\cite{Bench}.

Issues arise, however, when one considers specific supersymmetric grand
unified theories. One is the exact value of $\sin^2 \theta_W$, which
acquires important corrections from threshold effects at the electroweak
scale, associated with the spectrum of MSSM particles~\cite{EKNII,ELN},
and at the grand unification scale, associated with the spectrum of GUT
supermultiplets~\cite{EKNII,Heavy}.  Precision measurements indicate a 
small deviation of $\sin^2 \theta_W$ even from the value
predicted in a minimal supersymmetric $SU(5)$ GUT, assuming the range of
$\alpha_s(M_Z)$ now indicated by experiment~\cite{baggeretal}.

The second issue is the lifetime of the proton. Minimal supersymmetric
$SU(5)$ avoids the catastrophically rapid $p \to e^+ \pi^0$ decay that
scuppered non-supersymmetric $SU(5)$. However, supersymmetric $SU(5)$
predicts $p \to {\bar \nu} K^+$ decay through $d = 5$ operators at a rate
that may be too fast~\cite{PDK} to satisfy the presently available lower
limit on the lifetime for this decay~\cite{SK,PDG}. The latter requires
the $SU(5)$ colour-triplet Higgs particles to weigh $> 7.6 \times
10^{16}$GeV, whereas conventional $SU(5)$ unification for $\alpha_s(M_Z) =
0.1185 \pm 0.002$, $sin^2 \theta_W = 0.23117 \pm 0.00016$ and
$\alpha_{em}(M_Z) = 1/(127.943 \pm 0.027)$~\cite{PDG} would impose the
upper limit of $3.6 \times 10^{15}$~GeV at the 90\% confidence
level~\cite{PDK}. This problem becomes particularly acute if the sparticle
spectrum is relatively light, as would be indicated if the present
experimental and theoretical central values of $g_\mu - 2$~\cite{g-2}
remain unchanged as the errors are reduced.

The simplest way to avoid these potential pitfalls is to flip
$SU(5)$~\cite{f5,AEHN}.  As is well known, flipped $SU(5)$ offers the
possibility of decoupling somewhat the scales at which the Standard Model
$SU(3), SU(2)$ and $U(1)$ factors are unified. This would allow the
strength of the $U(1)$ gauge to become smaller than in minimal
supersymmetric $SU(5)$, for the same value of $\alpha_s(M_Z)$~\cite{ELN}.
Moreover, in addition to having a longer $p \to e/\mu^+ \pi^0$ lifetime 
than
non-supersymmetric $SU(5)$, flipped $SU(5)$ also suppresses the $d = 5$
operators that are dangerous in minimal supersymmetric $SU(5)$, by virtue
of its economical missing-partner mechanism~\cite{f5}.

In this paper, we re-analyze the issues of $\sin^2 \theta_W$ and proton
decay in flipped $SU(5)$~\cite{ELN}, in view of the most recent precise
measurements of $\alpha_s(M_Z)$ and $\sin^2\theta_W$, and the latest
limits on supersymmetric particles. We study these issues in the MSSM,
constraining the soft supersymmetry-breaking gaugino masses $m_{1/2}$ and
scalar masses $m_0$ to be universal at the GUT scale (CMSSM), making both
a general analysis in the $(m_{1/2}, m_0)$ plane and also more detailed
specific analyses of benchmark CMSSM parameter choices that respect all
the available experimental constraints~\cite{Bench}. We find that the $p
\to e/\mu^+ \pi^0$ decay lifetime exceeds the present experimental lower
limit~\cite{SK}, with a significant likelihood that it may be accessible
to the next round of experiments~\cite{UNO}. We recall the ambiguities and
characteristic ratios of proton decay modes in flipped $SU(5)$.

We first recall the lowest-order expression for $\alpha_s(M_Z)$ in 
conventional $SU(5)$ GUTs, namely
\begin{equation}
\alpha_s(M_Z)={{7 \over 3}\,\alpha\over 5\sin^2\theta_W-1}\ .
\label{eq:LO}
\end{equation}
The present central experimental value of
$\alpha_s(M_Z) = 0.118$ is obtained if one takes $\sin^2\theta_W=0.231$ 
and $\alpha^{-1}=128$, indicating the supersymmetric grand unification is 
in the right ball-park. However, at the next order, one should include
two-loop corrections $\delta_{\rm 2loop}$ as well as electroweak and GUT 
threshold 
corrections, that we denote by $\delta_{\rm light}$ and $\delta_{\rm 
heavy}$. Their effects can be included
by making the following substitution in (\ref{eq:LO})~\cite{EKNII}:
\begin{equation}
\sin^2\theta_W\to \sin^2\theta_W-\delta_{\rm 2loop}
-\delta_{\rm light}-\delta_{\rm heavy}\ ,
\label{eq:NLO}
\end{equation}   
where $\delta_{\rm 2loop}\approx0.0030$,
whereas $\delta_{\rm light}$ and
$\delta_{\rm heavy}$ can have either sign. If one neglects $\delta_{\rm 
light}$ and $\delta_{\rm heavy}$, the conventional $SU(5)$ prediction 
increases to $\alpha_s(M_Z)\approx0.130$~\cite{baggeretal}. 
A value of $\alpha_s(M_Z)$ within one standard deviation
of the present central experimental value requires 
$\delta_{\rm light}$ and/or $\delta_{\rm heavy}$ to be non-negligible,
so that the combination ($\delta_{\rm 2loop}
+\delta_{\rm light}+\delta_{\rm heavy}$) is suppressed. However, in large
regions of parameter space $\delta_{\rm light}>0$, which does not help. 
Moreover, in conventional $SU(5)$, as was pointed out 
in~\cite{EKNII,baggeretal},
a compensatory value of $\delta_{\rm heavy}$ is difficult 
to reconcile with proton decay constraints. This problem is exacerbated by 
the most recent lower limit on $\tau(p \to {\bar \nu} 
K^+)$~\cite{SK}~\footnote{It 
is true, as pointed out recently~\cite{GS}, that this is not a problem if 
one allows 
arbitrary squark mixing patterns. However, such options must respect 
low-energy flavour-changing neutral-interaction limits~\cite{EN}, and are 
not possible in the CMSSM.}.

As has been advertized previously~\cite{ELN}, an alternative way to lower 
$\alpha_s(M_Z)$ is to flip $SU(5)$.
In a flipped $SU(5)$ model, there is a first unification scale $M_{32}$ at
which the $SU(3)$ and $SU(2)$ gauge couplings become equal,
which is given to lowest order by~\cite{faspects}
\begin{eqnarray}
{1\over\alpha_2}-{1\over\alpha_5}&=&{b_2\over2\pi}\,
\ln{M_{32}\over M_Z}\ ,
\label{eq:RGE2}\\
{1\over\alpha_3}-{1\over\alpha_5}&=&{b_3\over2\pi}\,
\ln{M_{32}\over M_Z}\ ,
\label{eq:RGE3}
\end{eqnarray}
where $\alpha_2=\alpha/\sin^2\theta_W$, $\alpha_3=\alpha_s(M_Z)$, and the
one-loop beta function coefficients are $b_2=+1$, $b_3=-3$. The
hypercharge gauge coupling $\alpha_Y={5\over3}
(\alpha/\cos^2\theta_W)$ has, in general, a lower
value $\alpha_1'$ at the scale $M_{32}$:
\begin{equation}
{1\over\alpha_Y}-{1\over\alpha_1'}={b_Y\over2\pi}\, 
\ln{M_{32}\over M_Z}\ ,
\label{eq:RGEY}
\end{equation}
where $b_Y= 33 / 5$. Above the scale $M_{32}$, the gauge group is the 
full $SU(5) \times U(1)$, with the $U(1)$ gauge coupling $\alpha_1$ 
related to $\alpha_1'$ and the $SU(5)$ gauge coupling $\alpha_5$ as 
follows:
\begin{equation}
{25\over\alpha_1'}={1\over\alpha_5}+{24\over\alpha_1}\ .
\label{eq:U(1)}
\end{equation}
The $SU(5)$ and $U(1)$ gauge couplings then become
equal at some higher scale $M_{51}$. The maximum
possible value of $M_{32}$, namely $M^{\rm max}_{32}$, is obtained
by substituting $\alpha_1'=\alpha_5(M_{32})$ into (\ref{eq:RGEY}),
and coincides with the unification scale in conventional $SU(5)$:
$M^{\rm max}_{32} = M_Z \times exp((3 - 8\sin^2 \theta_W)\pi 
/14\alpha_{em}(M_Z))$, where $M_Z = 91.1882 \pm 0.0022$~GeV~\cite{PDG}. 
In general, one has
\begin{equation}
\alpha_s(M_Z)={{7 \over 3}\,\alpha\over 5\sin^2\theta_W-1
+{11\over2\pi}\,\alpha\ln(M^{\rm max}_{32}/M_{32})},
\label{eq:fLO} 
\end{equation}
and the flipped $SU(5)$ prediction for $\alpha_s(M_Z)$ is in general
smaller than in minimal SU(5), for the same value of $\sin^2 \theta_W$. 
The next-to-leading order corrections to (\ref{eq:fLO}) are also obtained
by the substitution in (\ref{eq:NLO}). Numerically, an increase
of $\sim10\%$ in the denominator in (\ref{eq:LO}),
which would compensate for
the decrease due to $\delta_{\rm 2loop}$, could be achieved simply by 
setting $M_{32}\approx{1\over3}M^{\rm max}_{32}$ in (\ref{eq:fLO}).   

\begin{figure}[h]
\begin{center}
\includegraphics[width=.7\textwidth,angle=0]{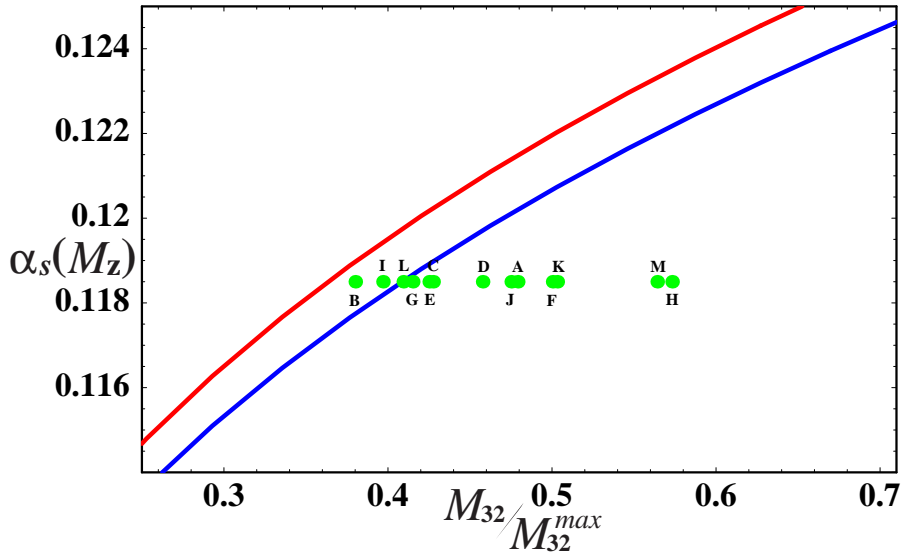}
\end{center}
\caption[]{\it
The solid lines show the correlation between $M_{32}$ in
flipped $SU(5)$ and $\alpha_s(M_Z)$ in the ${\overline {MS}}$
prescription, assuming $\sin^2 \theta_W^{\overline {MS}} = 0.23117 \pm
0.00016$, including $\delta_{\rm 2loop}$ but neglecting $\delta_{\rm
light}$ and $\delta_{\rm heavy}$. The points indicate the changes in $\tau
(p \to e/\mu^+ \pi^0)$ found for $\alpha_s(M_Z) = 0.1185$ and the central 
value of $\sin^2 \theta_W$ when including also the values of
$\delta_{\rm light}$ calculated for the CMSSM benchmark points.
}
\label{fig:m32alphas}
\end{figure}

In order to understand the implications for $\tau(p \to e/\mu^+ \pi^0)$
decay, we first calculate $M_{32}$, using (\ref{eq:fLO}) with
$\sin^2\theta_W$ replaced by $\sin^2\theta_W - \delta_{\rm 2loop}$,
leaving for later discussions of the possible effects of $\delta_{\rm
light, heavy}$.  Fig.~\ref{fig:m32alphas} exhibits the correlation between
$M_{32}$ and $\alpha_s(M_Z)$ in flipped $SU(5)$. The solid lines indicate
the range of values of $M_{32}$ allowed for a given value of
$\alpha_s(M_Z)$ (as given in the ${\overline {MS}}$ prescription),
assuming the experimentally-allowed range $\sin^2 \theta_W^{\overline
{MS}} = 0.23117 \pm 0.00016$~\cite{PDG}, and making no allowance for
either light or heavy thresholds. For the central experimental value
$\alpha_s(M_Z) = 0.1185$, we see immediately that $M_{32}$ is
significantly lower than its maximum value, which is $M^{\rm max}_{32} =
20.3 \times 10^{15}$~GeV for our central values of $\alpha_s(M_Z)$ and
$\sin^2\theta_W$.

We now explore the possible consequences of $\delta_{\rm light}$
for $M_{32}$, following~\cite{EKNII,ELN}. We approximate
the $\delta_{\rm light}$ correction by
\begin{eqnarray}
\delta_{\rm light}&=&{\alpha\over20\pi}\Bigl[
-3L(m_t)+{28 \over 3}L(m_{\tilde g})-{32 \over 3}L(m_{\tilde w})
-L(m_h)-4L(m_H)\nonumber\\  
&&+{5 \over 2}L(m_{\tilde q})-3L(m_{\tilde \ell_L})
+2L(m_{\tilde \ell_R})-{35 \over 36}L(m_{\tilde t_2})
-{19 \over 36}L(m_{\tilde t_1})\Bigr]\ ,
\label{eq:delta_l}
\end{eqnarray}
where $L(x)=\ln(x/M_Z)$. As already mentioned, we assume that the soft
supersymmetry-breaking scalar masses $m_0$, gaugino masses $m_{1/2}$ and
trilinear coefficients $A_0$ are universal at the GUT scale (CMSSM). We
used {\tt ISASUGRA}~\cite{ISASUGRA} to calculate the sparticle spectra in
terms of these quantities, $\tan \beta$ and the sign of $\mu$, assuming
$m_t = 175$~GeV~\footnote{Heavy singlet neutrinos were not used in the 
renormalization-group equations.}. In evaluating (\ref{eq:delta_l}), 
$m_{\tilde w}$ ($m_H$)
($m_{\tilde q}$) ($m_{\tilde \ell}$) were interpreted as the geometric 
means
of the chargino and neutralino ($H, A, H^\pm$) (${\tilde u}, {\tilde d},
{\tilde s}, {\tilde c}$) (${\tilde e}, {\tilde \mu}$) masses,
respectively, and the mixings of ${\tilde \tau}, {\tilde b}$ and ${\tilde
t}$ were all taken separately into account.

The unknown parameters in (\ref{eq:delta_l}) were constrained by requiring
that electroweak symmetry breaking be triggered by radiative corrections,
so that the correct overall electroweak scale and the ratio $\tan \beta$
of Higgs v.e.v.'s fix $|\mu|$ and $m_A$ in terms of $m_{1/2}$ and $m_0$.
Before making a more general survey, we recall that a number of benchmark
CMSSM scenarios have been proposed~\cite{Bench}, which include these
constraints and are consistent with all the experimental limits on
sparticle masses, the LEP lower limit on $m_h$, the world-average value of
$b \to s \gamma$ decay, the preferred range $0.1 < \Omega_\chi h^2 < 0.3$
of the supersymmetric relic density, and $g_\mu - 2$ within 2~$\sigma$ of
the present experimental value. These points all have $A_0 = 0$, but
otherwise span the possible ranges of $m_{1/2}, m_0, \tan \beta$ and
feature both signs for $\mu$.  Fig.~\ref{fig:m32alphas} also shows the
change in $M_{32}$ induced by the values of $\delta_{\rm light}$ in these
benchmark models, assuming a fixed value $\alpha_s(M_Z) = 0.1185$.  In
general, these benchmark models {\it increase} $M_{32}$ for any fixed
value of $\alpha_s(M_Z)$ and $\sin^2 \theta_W$.  As $\alpha_s(M_Z)$
varies, the predicted value of $M_{32}$ in each model varies in the same
way as indicated by the sloping lines. We recall that the estimated error
in $\alpha_s(M_Z)$ is about 0.002, corresponding to an uncertainty in
$M_{32}$ of the order of 20\%, and hence a corresponding uncertainty in
the proton lifetime of a factor of about two. The error associated with
the uncertainty in $\sin^2 \theta_W$ is somewhat smaller~\footnote{We note
from Fig.~\ref{fig:m32alphas} that {\it there is no benchmark model} for
which conventional $SU(5)$ grand unification is possible, with the
measured values of $\alpha_s(M_Z)$ and $\sin^2 \theta_W$, {\it unless} one
invokes GUT threshold effects.}.

We now turn to the calculation of $\tau(p \to e/\mu^+ \pi^0)$. We recall 
first 
that the form of the effective dimension-6 operator in flipped $SU(5)$ is 
different~\cite{faspects,ELNO} from that in conventional 
$SU(5)$~\cite{BEGN,EGN}:
\begin{eqnarray}
{\bar {\cal L}}_{\Delta B \ne 0} & =  & \frac{g_5^2}{2 M_{32}^2}   
 \left[ (\epsilon^{ijk}
{\bar d^c}_{k} e^{2 i \eta_{11}} \gamma^\mu P_L d_{j}) (u_{i}
 \gamma_\mu P_L \nu_L)  +  h.c. \right. \nonumber \\
 & + & \left. (\epsilon^{ijk}
({\bar d^c}_{k} e^{2 i \eta_{11}} \cos \theta_c
 + {\bar s^c}_{k} e^{2 i \eta_{21}} \sin \theta_c)
\gamma^\mu P_L u_{j}) (u_{i}
 \gamma_\mu P_L \ell_L) +  h.c. \right]
\label{32}
\end{eqnarray}
where $\theta_c$ is the Cabibbo angle~\footnote{Note the 
absence~\cite{faspects,ELNO} in the corresponding decay rate of the factor 
$(1 + (1 + |V_{ud}|^2)^2)$ found~\cite{BEGN,EGN} in conventional $SU(5)$, 
as recently re-emphasized in~\cite{PDK}. This {\it lengthens} $\tau_p$ 
by $\approx 5$ in flipped $SU(5)$, an effect that is typically
more than offset by the reduction in $M_{32}$.}. Also 
appearing in (\ref{32}) are 
two unknown but irrelevant CP-violating phases $\eta_{11,21}$ and lepton 
flavour eigenstates $\nu_L$ and $\ell_L$ that are related to mass 
eigenstates 
by unknown but relevant mixing matrices:
\beq
\nu_L =\nu_F U_\nu ~~~~~ , ~~\ell_L =\ell_F U_\ell.
\label{33}
\eeq
Despite our ignorance of the mixing matrices (\ref{33}), some 
characteristic flipped $SU(5)$ predictions can be made~\cite{faspects}:
\begin{eqnarray}
\Gamma(p \rightarrow e^+ \pi^o) = \frac{\cos ^2 \theta_c}{2}
 |U_{\ell_{11}}|^2
\Gamma(p \rightarrow {\bar \nu} \pi^+) = \cos ^2 \theta_c
|U_{\ell_{11}}|^2
\Gamma(n \rightarrow {\bar \nu} \pi^o) \nonumber \\
\Gamma(n \rightarrow e^+ \pi^-) = 2
 \Gamma(p \rightarrow e^+ \pi^o) ~~,~~
\Gamma(n \rightarrow \mu^+ \pi^-) = 2
\Gamma(p \rightarrow \mu^+ \pi^o) \nonumber \\
\Gamma(p \rightarrow \mu^+ \pi^o) = \frac{\cos ^2 \theta_c}{2}
 |U_{\ell_{12}}|^2
\Gamma(p \rightarrow {\bar \nu} \pi^+) = \cos ^2 \theta_c
|U_{\ell_{12}}|^2
\Gamma(n \rightarrow {\bar \nu} \pi^o)
\label{gammas}
\end{eqnarray}
In the light of recent experimental evidence for near-maximal 
neutrino mixing, it is reasonable to think that (at least some of) the 
$e/\mu$ entries in $U_\ell$ are ${\cal O}(1)$. 
In what follows, we assume that the lepton mixing factors 
$|U_{\ell_{11,12}}|^2$ are indeed ${\cal O}(1)$, and do not lead to
large numerical suppressions of both the $p \to e/\mu^+ \pi^0$ decay 
rates. 
Note that there is no corresponding suppression of the $p \to {\bar \nu} 
\pi^+$ and $n \to {\bar \nu} \pi^0$ decay rates, since all the neutrino 
flavours are summed over. However, without further information, we are 
unable to predict the ratio of $p \to e^+ X$ and $p \to \mu^+ X$ decay 
rates. Hereafter, 
wherever we refer to $p \to e^+ \pi^0$ decay, this mixing-angle ambiguity 
should be understood.

The $p \to e^+ \pi^0$ 
decay amplitude is proportional to the overall normalization of the 
proton wave function at the origin. The relevant matrix elements are 
$\alpha, \beta$, defined by
\begin{eqnarray}
\vev{0 | \epsilon_{ijk} (u^i d^j)_R u^k_L | p ({\mathbf k})} & \equiv & 
\alpha \, {\rm u}_L ({\mathbf k}), \\ 
\vev{0 | \epsilon_{ijk} (u^i d^j)_L u^k_L | p ({\mathbf k})} & \equiv & 
\beta \, {\rm u}_L ({\mathbf k}).
\label{alphabeta}
\end{eqnarray}
The reduced matrix elements $\alpha, \beta$ have recently been
re-evaluated in a lattice approach~\cite{lattice}, 
yielding values that are very similar and
somewhat larger than had often been assumed previously, and therefore 
exacerbating the proton-stability problem for conventional supersymmetric 
$SU(5)$. Here, we use
here the new central value $\alpha = \beta = 0.015$~GeV$^3$ for reference. The
error quoted on this determination is below 10\%, corresponding to an
uncertainty of less than 20\% in $\tau(p \to e^+ \pi^0)$, which would be
negligible compared with other uncertainties in our calculation. 
Thus, we have the following estimate, based on~\cite{ELNO,PDK} and 
references therein:
\begin{equation}
\tau(p \to e^+ \pi^0) \, = \, 3.8 \times 10^{35} \left( {M_{32} \over 
10^{16} 
{\rm GeV}} \right)^4 
\left( {\alpha_5(M^{\rm max}_{32}) \over \alpha_5(M_{32})} \right)^2 
\left( {0.015 {\rm 
GeV}^3 \over 
\alpha} \right)^2~{\rm y}
\label{eq:taup}
\end{equation}
for use in the subsequent analysis,
where we have absorbed reference values for $M_{32}$ and 
$\alpha_5(M_{32})$ as well as $\alpha$ and $\beta$, and 
$\alpha_5(M^{\rm max}_{32})/\alpha_5(M_{32}) = 
1 - (33/28) (\alpha_5(M^{\rm max}_{32})/2 \pi) {\rm 
ln}(M_{32}/M^{\rm max}_{32})$.

\begin{figure}
\vskip 0.5in
\vspace*{-0.75in}
\begin{minipage}{8in}
\epsfig{file=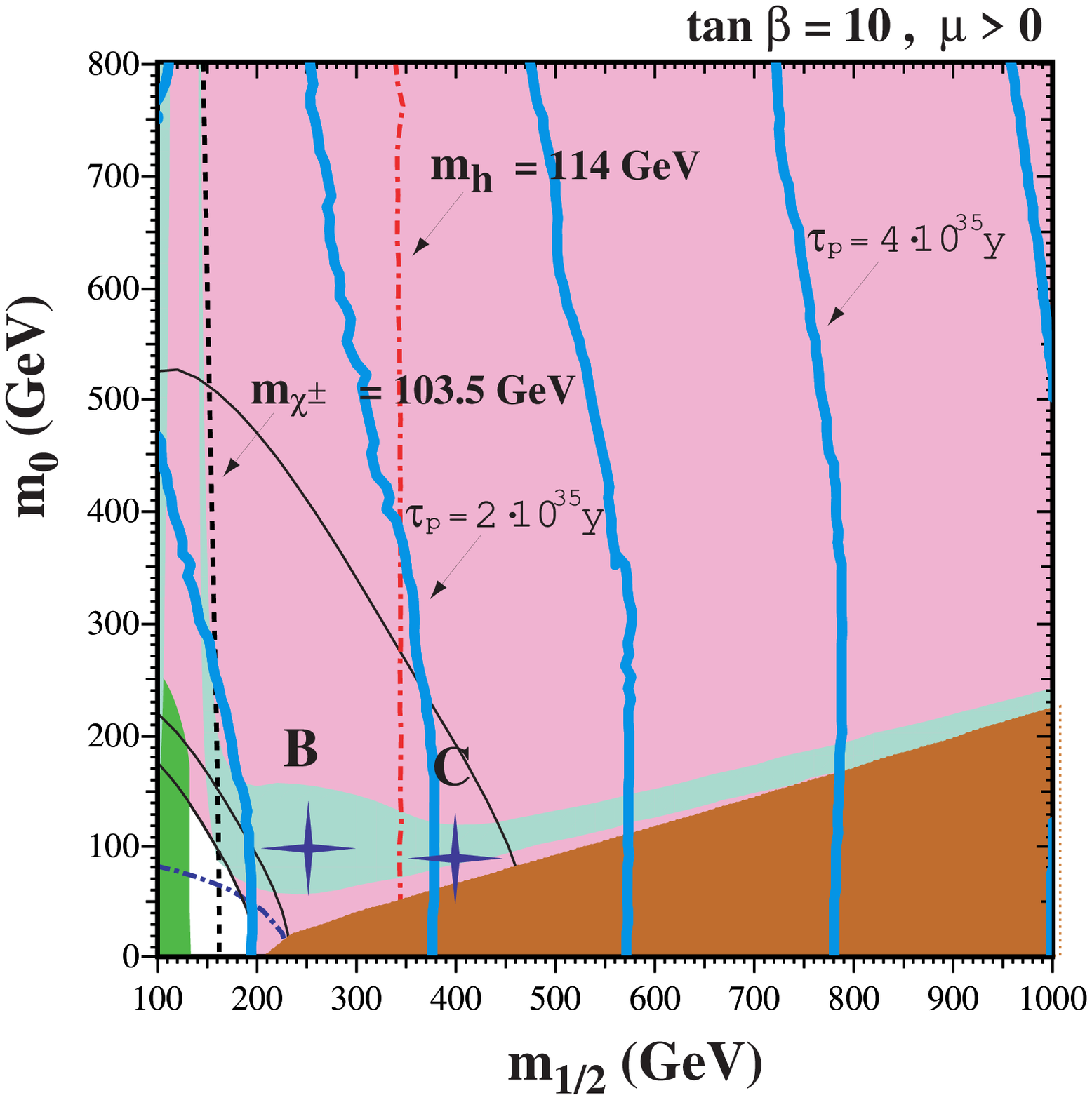,height=3.3in}
\hspace*{-0.17in}
\epsfig{file=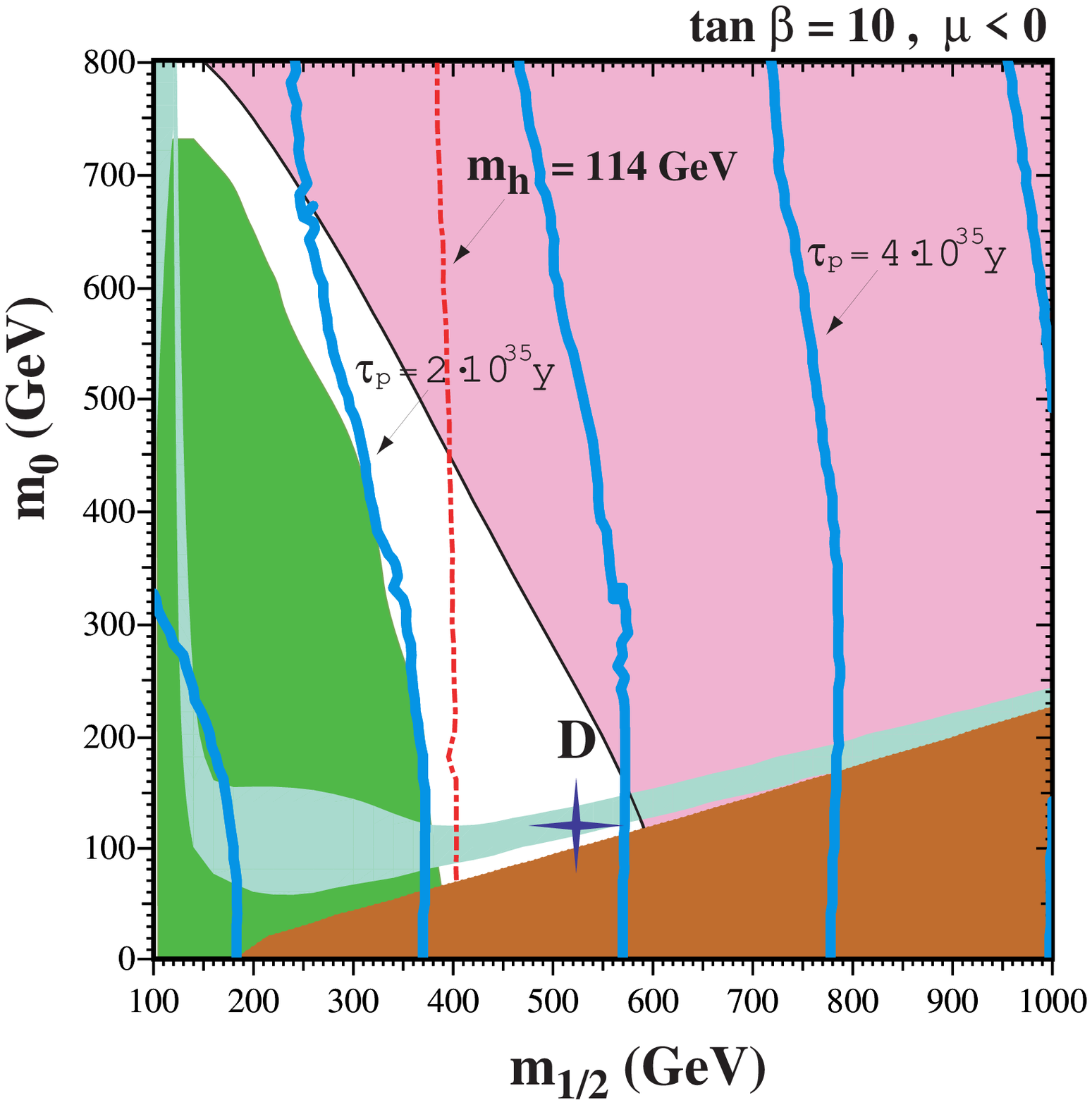,height=3.3in} \hfill
\end{minipage}
\begin{minipage}{8in}
\epsfig{file=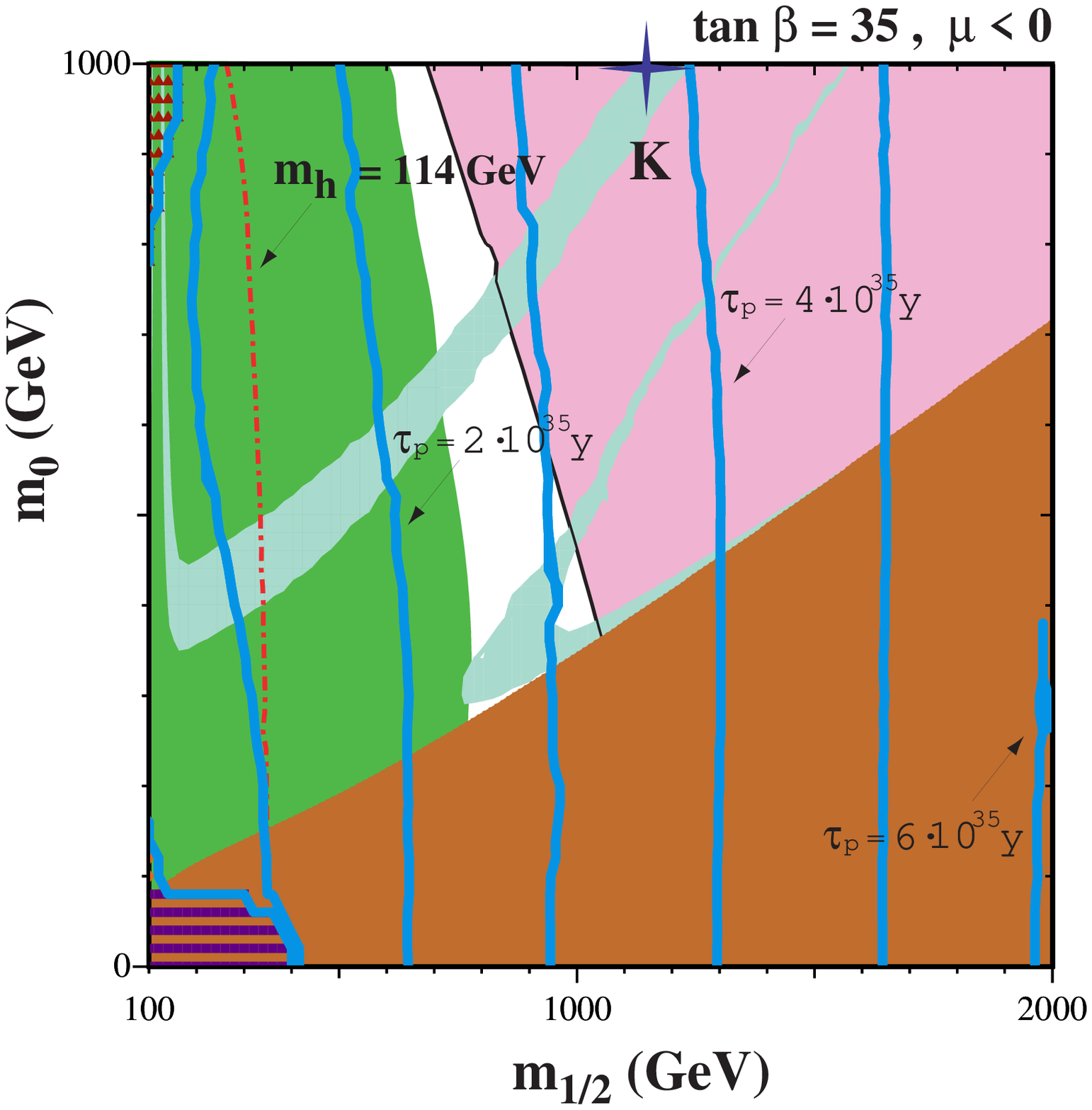,height=3.3in}
\hspace*{-0.2in}
\epsfig{file=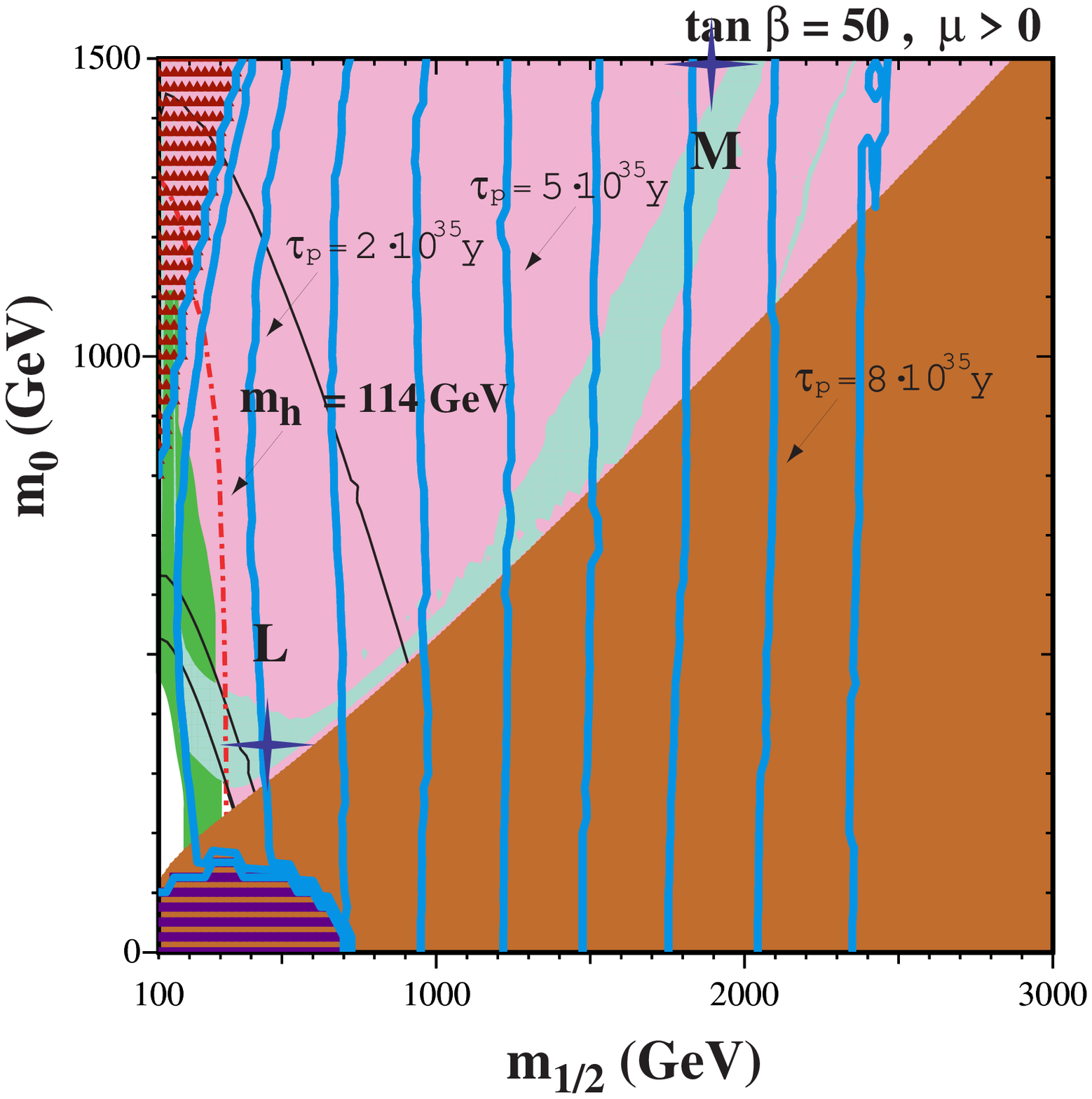,height=3.3in} \hfill
\end{minipage}
\caption{{\it 
The solid (blue) lines are contours of $\tau(p \to e/\mu^+ \pi^0)$ in the
$(m_{1/2}, m_0)$ plane for the CMSSM with (a) $\tan \beta = 10, \mu > 0$,
(b) $\tan \beta = 10, \mu < 0$, (c) $\tan \beta = 35, \mu < 0$ and (d)
$\tan \beta = 50, \mu > 0$. The (blue) crosses indicate the CMSSM
benchmark points with the corresponding value of $\tan \beta$ and sign of
$\mu$~\cite{Bench}. Following~\cite{EOS3}, the dark (red) shaded regions 
are excluded because the 
LSP is
charged, the light (turquoise) shaded regions have $0.1 < \Omega_\chi h^2
< 0.3$, intermediate (green) shaded regions at low $m_{1/2}$ are excluded
by $b \to s \gamma$, shaded (pink) regions at large $(m_{1/2}, m_0)$
are consistent with $g_\mu - 2$ at the 2-$\sigma$ level, and 
electroweak symmetry breaking is not possible in the hatched 
regions. The near-vertical
dashed (black) lines correspond to the LEP lower limit $m_\chi^\pm =
103.5$~GeV, the dot-dashed (red) lines to $m_h = 114$~GeV as calculated
using the {\tt FeynHiggs} code~\cite{FeynHiggs}, and the dotted (blue) 
lines at small
$(m_{1/2}, m_0)$ to $m_{\tilde e} = 100$~GeV.
}} 
\label{fig:planes} 
\end{figure}

We present a general view of flipped $SU(5)$ proton decay in the CMSSM in
Fig.~\ref{fig:planes}. The thick solid (blue) lines are contours of
$\tau(p \to e^+ \pi^0)$ for the indicated choices of $\tan \beta$ and the
sign of $\mu$~\footnote{The horizontal spacing between points sampled was 
comparable to the thickness of these lines.}, which span (most of) the 
range of possibilities. 
Where applicable, we have indicated by (blue) crosses and labels the CMSSM
benchmark points with the corresponding value of $\tan \beta$ and sign of
$\mu$~\cite{Bench}. Following~\cite{EOS3}, the dark (red) shaded regions
in the bottom right-hand parts of each panel are excluded because the LSP
is the lighter ${\tilde \tau}$:  astrophysics excludes a charged LSP. The
light (turquoise) shaded regions have LSP relic densities in the preferred
range $0.1 < \Omega_\chi h^2 < 0.3$ for cold dark matter. The intermediate
(green) shaded regions at lower $m_{1/2}$ are excluded by $b \to s
\gamma$, which is a more important constraint for $\mu < 0$.  The other
shaded (pink) regions at large $(m_{1/2}, m_0)$ are consistent with $g_\mu
- 2$ at the 2-$\sigma$ level. In panels (c) and (d), the hatched regions 
at low $m_{1/2}$ and
large $m_0$ are those where electroweak symmetry
breaking is no longer possible, and the horizontally-striped regions at 
low $m_0$ have tachyons. The dash-dotted (blue)  line at small
$(m_{1/2}, m_0)$ in panel (a) corresponds to $m_{\tilde e} = 100$~GeV. The
near-vertical dashed (black) lines at small $m_{1/2}$ correspond to the
LEP lower limit $m_\chi^\pm = 103.5$~GeV, and the dot-dashed (red) lines
to LEP lower limit $m_h = 114$~GeV as calculated using the {\tt FeynHiggs}
code~\cite{FeynHiggs}. In each case, only larger values of $m_{1/2}$ are
allowed, although there is uncertainty in the location of the $m_h$ 
line~\footnote{For fuller discussions of the implementations of these 
constraints with and without {\tt ISASUGRA}, see~\cite{Bench,EOS3}.}.

We see in Fig.~\ref{fig:planes} that the `bulk' regions of the parameter
space preferred by astrophysics and cosmology, which occur at relatively
small values of $(m_{1/2}, m_0)$, generally correspond to $\tau (p \to e^+
\pi^0) \sim (1~-~2) \times 10^{35}$~y. However, these `bulk' regions 
are generally disfavoured by the experimental lower limit on $m_h$ 
and/or by $b \to s \gamma$ decay. Larger values of $\tau (p
\to e^+ \pi^0)$ are found in the `tail' regions of the cosmological
parameter space, which occur at large $m_{1/2}$ where $\chi - {\tilde
\ell}$ coannihilation may be important, and at larger $m_{1/2}$ and $m_0$
where resonant direct-channel annihilation via the heavier Higgs bosons
$A, H$ may be important.

We turn finally to the possible implications of the GUT threshold effect 
$\delta_{\rm heavy}$~\cite{EKNII,Heavy}. A general expression for this in 
flipped $SU(5)$ is given in~\cite{EKNII}:
\begin{equation}
\delta_{\rm heavy}={\alpha\over20\pi}
\left[ -6\ln{M_{32}\over M_{H_3}}-6\ln{M_{32}\over M_{\bar H_3}}
+4\ln{M_{32}\over M_V}\right]={\alpha\over20\pi}
\left[ -6\ln{r^{4/3}g^{2/3}_5\over\lambda_4\lambda_5}\right]
\label{heavyfSU5}
\end{equation}
where $M_{H_3}=\lambda_4|V|$ and $M_{\bar H_3}=\lambda_5|V|$ are the 
masses of
the heavy triplet Higgs supermultiplets, the $X,Y$ gauge bosons and
gauginos have common masses $M_V=g_5|V|$ where $V$ is the common v.e.v. 
of the
${\mathbf 10}$ and ${\mathbf {\overline {10}}}$ Higgs supermultiplets, 
$\lambda_{4,5}$ 
are (largely
unconstrained) Yukawa couplings, $g_5$ is the $SU(5)$ gauge coupling, and
$r \equiv {\rm max}\{g_5,\lambda_4,\lambda_5\}$. Thanks to the economical
missing-partner mechanism of flipped $SU(5)$, the $H_3$ and $\bar H_3$ do
not mix, and hence do not contribute significantly to proton decay. Thus
there is no strong constraint on $M_{H_3,\bar H_3}$ from proton decay in
flipped $SU(5)$, and it is possible that $M_{H_3,\bar H_3}<M_V$ (\ie,
$r=g_5$). In this case, we can see from (\ref{heavyfSU5}) that
$\delta_{\rm heavy} < 0$ naturally. For instance, as pointed out
in~\cite{ELN}, if $\lambda_4,\lambda_5\sim{1\over8}g_5$, then $\delta_{\rm
heavy}\approx-0.0030$, which completely compensates the $\delta_{\rm
2loop}$ contribution. 

\begin{figure}[h]
\begin{center}  
\includegraphics[width=.7\textwidth,angle=0]{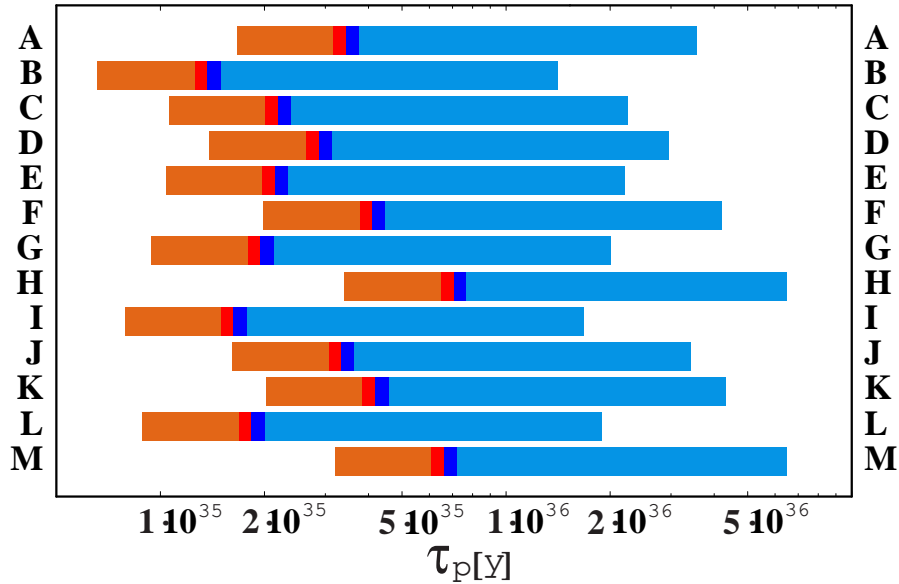}
\end{center}
\caption[]{\it
For each of the CMSSM benchmark points, this plot shows,
by the lighter
outer bars, the range of $\tau (p \to e/\mu^+ \pi^0)$ attained by varying
$\delta_{\rm heavy}$ over the range -0.0016 to + 0.0005~\cite{ELN}.  The 
central boundary of
the narrow inner bars (red, blue) corresponds to the effect of
$\delta_{\rm light}$ alone, with $\delta_{\rm heavy} = 0$, while the
narrow bars themselves represent uncertainty in $\sin^2 \theta_W$.
We see that heavy threshold effects could make $\tau (p \to e/\mu^+ 
\pi^0)$    
slightly shorter or considerably longer.
} 
\label{fig:lines} \end{figure}

We also recall that, in general,
including $\delta_{\rm heavy}$ leads to a re-scaling
of the $M_{32}/M^{\rm max}_{32}$:
\begin{equation}
{M_{32}\over M^{\rm max}_{32}}\to {M_{32}\over M^{\rm max}_{32}}
\ e^{-10\pi\,\delta_{\rm heavy}/11\alpha}\ .
\label{scaling}
\end{equation}
We display in Fig.~\ref{fig:lines} the possible numerical effects of
$\delta_{\rm heavy}$ on $\tau (p \to e/\mu^+ \pi^0)$ in the various 
benchmark
scenarios, assuming the plausible ranges $-0.0016 < \delta_{\rm heavy} < 
0.0005$~\cite{ELN}. The boundary between the different shadings for each 
strip
corresponds to the case where $\delta_{\rm heavy} = 0$. The left (red)  
parts of the strips show how much $\tau (p \to e^+ \pi^0)$ could be
reduced by a judicious choice of $\delta_{\rm heavy}$, and the right
(blue) parts of the strips show how much $\tau (p \to e^+ \pi^0)$ 
could be
increased. The inner bars correspond to the uncertainty in $\sin^2 
\theta_W$. On the optimistic side, we see that some models could yield 
$\tau (p \to e^+ \pi^0) < 10^{35}$~y, and all models might have $\tau (p 
\to e^+ \pi^0) < 5 \times 10^{35}$~y. However, on the pessimistic side, in 
no model can we exclude the possibility that $\tau (p \to e^+ \pi^0) > 
10^{36}$~y.

We recall that a new generation of massive water-{\v C}erenkov detectors
weighing up to $10^6$~tonnes is being proposed~\cite{UNO}, that may be
sensitive to $\tau (p \to e^+ \pi^0) < 10^{35}$~y. According to our
calculations, such an experiment has a chance of detecting proton decay in
flipped $SU(5)$, though nothing can of course be guaranteed. We recall
that there is a mixing-angle ambiguity (\ref{gammas}) in the final-state
charged lepton, so any such next-generation detector should be equipped to
detect $e^+$ and/or $\mu^+$ equally well. We also
recall~\cite{faspects,ELNO} that flipped $SU(5)$ makes predictions
(\ref{gammas}) for ratios of decay rates involving strange particles,
neutrinos and charged leptons that differ characteristically from those of
conventional $SU(5)$. Comparing the rates for $e^+$, $\mu^+$ and neutrino
modes would give novel insights into GUTs as well as mixing patterns.

We conclude that flipped $SU(5)$ evades two of the pitfalls of
conventional supersymmetric $SU(5)$. As we have shown in this paper, it
offers the possibility of lowering the prediction for $\alpha_s(M_Z)$ for
any given value of $\sin^2 \theta_W$ and choice of sparticle spectrum. As
for proton decay, we first recall that flipped $SU(5)$ suppresses $p \to
{\bar \nu} K^+$ decay naturally via its economical missing-partner
mechanism. As in conventional supersymmetric $SU(5)$, the lifetime for $p
\to e/\mu^+ \pi^0$ decay generally exceeds the present experimental lower
limit. However, as we have shown in this paper, the flipped $SU(5)$
mechanism for reducing $\alpha_s(M_Z)$ reduces the scale $M_{32}$ at which
colour $SU(3)$ and electroweak $SU(2)$ are unified, bringing $\tau(p \to
e/\mu^+ \pi^0)$ tantalizingly close to the prospective sensitivity of the
next round of experiments. Proton decay has historically been an
embarrassment for minimal $SU(5)$ GUTs, first in their non-supersymmetric
guise and more recently in their minimal supersymmetric version. The
answer may be to flip $SU(5)$ out of trouble.

\section*{Acknowledgements}
The work of D.V.N. was partially supported by DOE grant
DE-F-G03-95-ER-40917.


\begin{thebibliography}{99}

\bibitem{EKN}
J.~R.~Ellis, S.~Kelley and D.~V.~Nanopoulos,
Phys.\ Lett.\ B {\bf 249} (1990) 441;
Phys.\ Lett.\ B {\bf 260} (1991) 131;
U.~Amaldi, W.~de Boer and H.~Furstenau,
Phys.\ Lett.\ B {\bf 260} (1991) 447;
C.~Giunti, C.~W.~Kim and U.~W.~Lee,
Mod.\ Phys.\ Lett.\ A {\bf 6} (1991) 1745.

\bibitem{Ibanez}
L.~E.~Ibanez,
arXiv:hep-ph/0109082.

\bibitem{power}
K.~R.~Dienes, E.~Dudas and T.~Gherghetta,
Phys.\ Lett.\ B {\bf 436} (1998) 55
[arXiv:hep-ph/9803466];
Nucl.\ Phys.\ B {\bf 537} (1999) 47
[arXiv:hep-ph/9806292].

\bibitem{Georgi}
N.~Arkani-Hamed, A.~G.~Cohen and H.~Georgi,
arXiv:hep-th/0108089.

\bibitem{Ross}
D.~M.~Ghilencea and G.~G.~Ross,
Nucl.\ Phys.\ B {\bf 606} (2001) 101
[arXiv:hep-ph/0102306].

\bibitem{LEPEWWG}
LEP Collaborations, LEP Electroweak Working Group and SLD Heavy-Flavour 
Working Group, LEPEWWG/2002-01, available from \\
{\tt http://lepewwg.web.cern.ch/LEPEWWG/stanmod/}.

\bibitem{SusyHiggs}
Y.~Okada, M.~Yamaguchi and T.~Yanagida,
Prog.\ Theor.\ Phys.\  {\bf 85} (1991) 1;
J.~R.~Ellis, G.~Ridolfi and F.~Zwirner,
Phys.\ Lett.\ B {\bf 257} (1991) 83;
H.~E.~Haber and R.~Hempfling,
Phys.\ Rev.\ Lett.\  {\bf 66} (1991) 1815.

\bibitem{EHNOS}
J. Ellis, J.S. Hagelin, D.V. Nanopoulos, K.A. Olive
and M. Srednicki, Nucl. Phys. B {\bf 238} (1984) 453; see also
H. Goldberg, Phys. Rev. Lett. {\bf 50} (1983) 1419.

\bibitem{bsg}
M.S. Alam et al., [CLEO Collaboration], Phys.\ Rev.\ Lett.\ {\bf 74}
(1995) 2885 as updated in
S.~Ahmed et al., {CLEO CONF 99-10};
BELLE Collaboration, BELLE-CONF-0003, contribution to the 30th
International conference on High-Energy Physics, Osaka, 2000.
See also
K.~Abe {\it et al.},  [Belle Collaboration],
[arXiv:hep-ex/0107065];
L.~Lista  [BaBar Collaboration],
[arXiv:hep-ex/0110010];
C. Degrassi, P. Gambino and G.~F. Giudice,
JHEP {\bf 0012} (2000) 009 [arXiv:hep-ph/0009337];
M.~Carena, D.~Garcia, U.~Nierste and C.~E.~Wagner,
Phys. Lett. B {\bf 499} (2001) 141
[arXiv:hep-ph/0010003].
D.~A.~Demir and K.~A.~Olive,
Phys.\ Rev.\ D {\bf 65} (2002) 034007
[arXiv:hep-ph/0107329];
P.~Gambino and M.~Misiak,
Nucl.\ Phys.\ B {\bf 611} (2001) 338
[arXiv:hep-ph/0104034].

\bibitem{g-2}
H.~N.~Brown {\it et al.}  [Muon g-2 Collaboration],
Phys.\ Rev.\ Lett.\  {\bf 86} (2001) 2227
[arXiv:hep-ex/0102017].

\bibitem{Bench}
M.~Battaglia {\it et al.}, Eur. Phys. J. C {\bf 22} (2001) 535
[arXiv:hep-ph/0106204].

\bibitem{EKNII}
J.~R.~Ellis, S.~Kelley and D.~V.~Nanopoulos,
Nucl.\ Phys.\ B {\bf 373} (1992) 55;
Phys.\ Lett.\ B {\bf 287} (1992) 95
[arXiv:hep-ph/9206203].

\bibitem{ELN}
J.~R.~Ellis, J.~L.~Lopez and D.~V.~Nanopoulos,
Phys.\ Lett.\ B {\bf 371} (1996) 65
[arXiv:hep-ph/9510246].

\bibitem{Heavy}
R. Barbieri and L. Hall, Phys. Rev. Lett. {\bf 68} (1992) 752; 
J. Hisano, H. Murayama, and T. Yanagida, Phys. Rev. 
Lett. {\bf 69} (1992) 1014; 
K. Hagiwara and Y. Yamada, Phys. Rev. Lett. {\bf 70} (1993) 709; 
F. Anselmo, L. Cifarelli, A. Peterman, and A. Zichichi,
Nuovo Cim. {\bf105A} (1992) 1025; 
P. Langacker and N. Polonsky, Phys. Rev.{\bf 47} (1993) 4028.

\bibitem{baggeretal}
M. Bastero-Gil and J. Perez-Mercader, Phys. Lett. B 
{\bf 322} (1994) 355; 
A. Faraggi and B. Grinstein, Nucl. Phys. B {\bf 422} (1994) 3;
P. Chankowski, Z. Pluciennik and S. Pokorski, Nucl. 
Phys. B {\bf 439} (1995) 23; 
P. Langacker and N. Polonsky, arXiv:hep-ph/9503214;
L. Clavelli and P. Coulter, Phys. Rev. D {\bf 51} (1995) 3913; 
J. Bagger, K. Matchev, and D. Pierce, Phys. Lett. B {\bf 348} (1995) 443.   

\bibitem{PDK}
H.~Murayama and A.~Pierce,
Phys.\ Rev.\ D {\bf 65} (2002) 055009
[arXiv:hep-ph/0108104].

\bibitem{SK}
M.~Shiozawa {\it et al.}  [Super-Kamiokande Collaboration],
Phys.\ Rev.\ Lett.\  {\bf 81} (1998) 3319
[arXiv:hep-ex/9806014];
Y.~Hayato {\it et al.}  [SuperKamiokande Collaboration],
Phys.\ Rev.\ Lett.\  {\bf 83} (1999) 1529
[arXiv:hep-ex/9904020].

\bibitem{PDG}
D.~E.~Groom {\it et al.}  [Particle Data Group Collaboration],
Eur.\ Phys.\ J.\ C {\bf 15} (2000) 1.

\bibitem{f5}
S.~M.~Barr,
Phys.\ Lett.\ B {\bf 112} (1982) 219;
J.~P.~Derendinger, J.~E.~Kim and D.~V.~Nanopoulos,
Phys.\ Lett.\ B {\bf 139} (1984) 170;
I.~Antoniadis, J.~R.~Ellis, J.~S.~Hagelin and D.~V.~Nanopoulos,
Phys.\ Lett.\ B {\bf 194} (1987) 231.

\bibitem{AEHN}
I.~Antoniadis, J.~R.~Ellis, J.~S.~Hagelin and D.~V.~Nanopoulos,
Phys.\ Lett.\ B {\bf 205} (1988) 459;
Phys.\ Lett.\ B {\bf 208} (1988) 209
[Addendum-ibid.\ B {\bf 213} (1988) 562];
Phys.\ Lett.\ B {\bf 231} (1989) 65.

\bibitem{UNO}
C.~K.~Jung,
arXiv:hep-ex/0005046.

\bibitem{GS}
B.~Bajc, P.~F.~Perez and G.~Senjanovic,
arXiv:hep-ph/0204311.

\bibitem{EN}
J.~R.~Ellis and D.~V.~Nanopoulos,
Phys.\ Lett.\ B {\bf 110} (1982) 44;
R.~Barbieri and R.~Gatto,   
Phys.\ Lett.\ B {\bf 110} (1982) 211.

\bibitem{faspects}
J.~R.~Ellis, J.~S.~Hagelin, S.~Kelley and D.~V.~Nanopoulos,
Nucl.\ Phys.\ B {\bf 311} (1988) 1.

\bibitem{ISASUGRA}
We use version {\tt 7.51} of
H.~Baer, F.~E.~Paige, S.~D.~Protopopescu and X.~Tata,
{\it ISAJET 7.48: A Monte Carlo event generator for $p p, {\bar p} p$, and
$e^+ e^-$ reactions}, [arXiv:hep-ph/0001086],
modified as in version {\tt 7.58}.

\bibitem{ELNO}
J.~R.~Ellis, J.~L.~Lopez, D.~V.~Nanopoulos and K.~A.~Olive,
Phys.\ Lett.\ B {\bf 308} (1993) 70
[arXiv:hep-ph/9303307].

\bibitem{BEGN}
A.~J.~Buras, J.~R.~Ellis, M.~K.~Gaillard and D.~V.~Nanopoulos,
Nucl.\ Phys.\ B {\bf 135} (1978) 66.

\bibitem{EGN}
J.~R.~Ellis, M.~K.~Gaillard and D.~V.~Nanopoulos,
Phys.\ Lett.\ B {\bf 88} (1979) 320.

\bibitem{lattice}
Y.~Kuramashi  [JLQCD Collaboration],
arXiv:hep-ph/0103264.

\bibitem{EOS3}
J.~R.~Ellis, K.~A.~Olive and Y.~Santoso,
arXiv:hep-ph/0202110.

\bibitem{FeynHiggs}
S.~Heinemeyer, W.~Hollik and G.~Weiglein,
Comput.\ Phys.\ Commun.\  {\bf 124} (2000) 76
[arXiv:hep-ph/9812320];
S.~Heinemeyer, W.~Hollik and G.~Weiglein,
Eur.\ Phys.\ J.\ C {\bf 9} (1999) 343
[arXiv:hep-ph/9812472].

\end{thebibliography}
\end{document}